\documentclass[fleqn,twoside]{article}
\usepackage{espcrc2}

\usepackage{graphicx}
\usepackage[figuresright]{rotating}

\newcommand{\AmS}{{\protect\the\textfont2
  A\kern-.1667em\lower.5ex\hbox{M}\kern-.125emS}}

\def\gtsim{\footnotesize{\mathop{\raisebox{-.4ex}{\rlap{$\sim$}}
\raisebox{.4ex}{$>$}}}}

\title{Sensitivity of the NEMO telescope to neutrinos from microquasars}

\author{C. Distefano\address[LNS]{LNS-INFN, via S. Sofia 62, 95123 Catania, Italy}
        for the NEMO Collaboration}

\begin{document}

\begin{abstract}
We present the results of Monte Carlo simulation studies of the capability of
the proposed NEMO telescope to detect TeV muon neutrinos from Galactic microquasars.
In particular we determined the number of the detectable events from each known
microquasar together with the expected atmospheric neutrino and muon background events.
We also discuss the detector sensitivity to neutrino fluxes expected from
microquasars, optimizing the event selection in order to reject the
atmospheric background, and we show the number of events surviving the
event selection.
\vspace{1pc}
\end{abstract}

\maketitle

\section{INTRODUCTION}

Microquasars are Galactic X-ray binary systems which exhibit
relativistic jets, observed in the radio band \cite{Chaty05}.
Several authors propose microquasar jets as sites of acceleration of charged particles
up to energies of about $10^{16}$ eV, and of high energy neutrino production.
According to present models, neutrinos could be produced both in p$\gamma$ \cite{Levinson01,Distefano02}
and pp interaction scenarions \cite{Bednarek05,Aharonian05,Christiansen05,Romero05}.

The aim of this paper is to study the possibility to detect neutrinos from
known microquasars with the proposed NEMO-km$^3$ telescope \cite{nemo}.
In particular, for each microquasar we simulated the number of expected
events according to present models. We also simulated the expected
background from atmospheric neutrinos and muons.

We applied an event selection in order to reject the atmospheric background.
The event selection is a combination of
different criteria, chosen optimizing the detector sensitivity
to neutrino fluxes coming from each microquasar.
Eventually we calculated the expected number of events surviving
the selection referring to 1 year of data taking.

\section{DETECTOR LAY-OUT}

The NEMO-km$^3$ telescope, simulated in this paper, is a square array of
$9\times9$ towers with a distance between towers of 140 m.
In this configuration each tower hosts 72 PMTs (with a diameter of 10"), namely 5832 PMTs for the
whole detector with a total geometrical volume of $\sim0.9$ km$^3$.
We considered an 18 storey tower; each storey
is made of a 20 m long beam structure hosting two optical modules (one downlooking and one looking horizontally) at each end
(4 OMs per storey). The vertical distance between storeys is 40 m. A spacing of 150 m is added at the base of the tower,
between the anchor and the lowermost storey.

The detector response is simulated
using the simulation codes developed by the ANTARES
Collaboration \cite{antares_codes}. In the simulation codes, the light
absorption length, measured in the site of Capo Passero
($L_a\approx 68$ m at 440 nm \cite{Riccobene05}), is taken into account. Once
the sample of PMT hits is generated, spurious PMT hits,
due to the underwater optical noise ($^{40}$K decay), are
introduced, with a rate of 30 kHz for 10"
PMTs, corresponding to the average value measured in Capo
Passero site.

\section{EXPECTED NEUTRINO EVENTS FROM MICROQUASARS}

For each microquasar, a number $N_{total}=10^9$ of interacting
neutrinos was simulated in the energy range 1 - 100 TeV
in order to reproduce the neutrino fluxes calculated by Distefano et al.
\cite{Distefano02}. The generation spectral index is chosen to be $X=1$, in order to guarantee
a good event statistics at the highest energies.
The simulated events are then weighted to the theoretical fluxes
and the number of expected events is computed and reported in
Tab. \ref{tab:N_mu_simulated}.
Results refer to an integration time $\Delta t$ equal to
the duration of the considered burst
for the transient sources and to one year for the steady
sources.

For microquasar LS 5039, $N_{total}=10^8$ events were generated
in the energy range 0.1 - 1 TeV to simulate neutrinos events
according to the model proposed by Aharonian et al. \cite{Aharonian05}.
In Tab. \ref{tab:N_mu_aharonian} the number of muon events
from LS 5039 is reported.
The simulations refer to a power-law neutrino spectrum,
$dn_\nu/d\varepsilon_\nu\propto\varepsilon_\nu^{-\Gamma}$,
with energy cutoff $\varepsilon_\nu^{max}=10$ TeV and 100 TeV,
and $\Gamma=1.5$ and 2, respectively. For the four combinations of parameters
$\Gamma$ and $\varepsilon_\nu^{max}$, an average neutrino energy flux
$f_\nu^{th}=10^{-10}$ erg/cm$^2$ s ($\varepsilon_\nu>0.1$ TeV) was simulated.

For microquasar Cygnus X-3, $N_{total}=10^8$ events were generated
in the energy range 0.1 - 1 TeV and $N_{total}=10^9$ events with energies 100 - 1000 TeV
to simulate Bednarek's theoretical fluxes \cite{Bednarek05}.
In Tab. \ref{tab:N_mu_bednarek} we give the number of muon events
from Cygnus X-3, considering different values of the the spectral index $\kappa$ and for the
Lorentz factor cut-off $\gamma_A^{max}$ of the spectrum of nuclei generating
the neutrinos. The case of neutrons injected by mono-energetic nuclei having a Lorentz factor
$\gamma_A^{min}\approx10^5$ is also considered (see \cite{Bednarek05} for details).

Other models, present in literature, are not considered here because
they refer to microquasars outside the field of view of the NEMO telescope \cite{Christiansen05}
or because the predicted neutrino fluxes are too low to be detected by a km$^3$ telescope \cite{Romero05}.

\begin{table*}
\caption{Expected number of neutrino induced muons from microquasars
expected from the theoretical neutrino energy flux
quoted by Distefano et al. \cite{Distefano02}:
$N_\mu^{exp}$ is the total number of reconstructed muons from each
microquasar during the time interval $\Delta t$, $N_\mu^{m}$ is the expected number of selected muon neutrino events
from the sources and $N_\mu^{b}$ is the atmospheric background events
surviving the event selection and expected in 1 year of data taking.}
\label{tab:N_mu_simulated}
\newcommand{\m}{\hphantom{$-$}}
\newcommand{\cc}[1]{\multicolumn{1}{c}{#1}}
\renewcommand{\tabcolsep}{2pc} 
\renewcommand{\arraystretch}{1.2} 
\begin{tabular}{@{}lcc|cc}
\hline
Steady Microquasars       &  $\Delta t$ (days) &  $N_\mu^{exp}$   & $N_\mu^{m}$ &  $N_\mu^{b}$  \\
\hline
  LS 5039                 & 365                       & 0.2       &     0.1     &   0.1      \\
  Scorpius X-1            & 365                       & 0.8       &     0.2     &   0.1      \\
  SS433                   & 365                       & 205.0     & 76.0        &   0.1              \\
  GX 339-4                & 365                       & 185.0     &     68.0        &   0.1              \\
  Cygnus X-1              & 365                       & 2.0       &     0.5         &   0.1      \\
\hline
Bursting Microquasars \\
\hline
  XTE J1748-288           & 20                        & 2.2       &     0.8     &   0.3             \\
  Cygnus X-3              & 3                         & 3.4       &     0.8     &   0.1     \\
  GRO J1655-40            & 6                         & 1.7       &     0.6     &   0.1             \\
  GRS 1915+105            & 6                         & 0.4       &     0.1     &   $<0.1$          \\
  Circinus X-1            & 4                         & 0.2       &     0.1     &   0.1             \\
  XTE J1550-564           & 5                         & $<0.1$    &     $<0.1$      &   $<0.1$      \\
  V4641 Sgr               & 0.3                       & $<0.1\div4.6$    &  $<0.1\div$1.4   &   0.1     \\
  GS 1354-64              & 2.8                       & $<0.1$    &     $<0.1$      &   0.1         \\
  GRO J0422+32            & 1$\div$20                 & 0.1$\div$1.5    &   $<0.1\div$0.4   &   0.1     \\
  XTE J1118+480           & 30$\div$150               & 2.4$\div$12.0   &   1.0$\div$4.8    &   0.2     \\
\hline
\end{tabular}\\[2pt]
\end{table*}

\begin{table*}
\caption{Expected number $N_\mu^{exp}$ of neutrino induced muons from microquasar LS 5039 according to Aharonian et al. model \cite{Aharonian05},
during an time interval of 1 year. We also give the number $N_\mu^{m}$ from the source compared to the atmospheric background events $N_\mu^{b}$
surviving the event selection.}
\label{tab:N_mu_aharonian}
\newcommand{\m}{\hphantom{$-$}}
\newcommand{\cc}[1]{\multicolumn{1}{c}{#1}}
\renewcommand{\tabcolsep}{2pc} 
\renewcommand{\arraystretch}{1.2} 
\begin{tabular}{@{}ccc|cc}
\hline
$\Gamma$ & $\varepsilon_\nu^{max}$ (TeV) &  $N_\mu^{exp}$  &    $N_\mu^{m}$ &  $N_\mu^{b}$\\
\hline
  1.5            & ~10      & 7.0   & 1.7      & 0.2      \\
  1.5            & 100      & 15.0  & 4.9      & 0.1      \\
  2.0            & ~10      & 4.3   & 1.0      & 0.3      \\
  2.0            & 100      & 9.0   & 2.6      & 0.1      \\
\hline
\end{tabular}\\[2pt]
\end{table*}

\begin{table*}
\caption{Expected number of neutrino induced events from microquasar Cygnus X-3 (Bednarek model \cite{Bednarek05}):
$N_\mu^{exp}$ is the total number of reconstructed muons during an time interval of 1 year; $N_\mu^{m}$ and
$N_\mu^{b}$ are the number of source and background events surviving the selection.}
\label{tab:N_mu_bednarek}
\newcommand{\m}{\hphantom{$-$}}
\newcommand{\cc}[1]{\multicolumn{1}{c}{#1}}
\renewcommand{\tabcolsep}{2pc} 
\renewcommand{\arraystretch}{1.2} 
\begin{tabular}{@{}ccc|ccc}
\hline
Cygnus X-3 &    REG. II   &  REG. III & REG. II   &  REG. III &   \\
\hline
     $~~~\kappa,~\gamma_A^{max}$             & $N_\mu^{exp}$   & $N_\mu^{exp}$  & $N_\mu^{m}$   & $N_\mu^{m}$  &  $N_\mu^{b}$ \\
\hline
  $2.0,~10^6$            & 2.0             &  1.8     & 0.4            &  0.4      &  0.1      \\
  $2.0,~10^7$            & 5.2             &  5.0     & 1.1            &  1.1      &  0.1      \\
  $2.5,~10^7$            & 0.3             &  0.2     & $<0.1$             &  0.1      &  0.1      \\
mono-energetic           & $<0.1$          &  4.5     & $<0.1$             &  0.7      &  0.1      \\
\hline
\end{tabular}\\[2pt]
\end{table*}

\subsection{SIMULATION OF THE BACKGROUND}

A sample of $7\cdot10^9$ atmospheric neutrinos have been generated
using the ANTARES event generation code, based on a weighted
generation technique \cite{antares_codes}. The events were generated in the energy
range $10^2\div10^8$ GeV, with a spectral index $X=2$ and a $4\pi$
isotropic angular distribution. The events were then weighted
to to the sum of the Bartol flux \cite{Agrawal96} and of prompt
neutrino {\tt rqpm} model \cite{Bugaev98} flux. So doing, we
compute a number $\approx 4\cdot 10^4$ of detected atmospheric
neutrino events per year of data acquisition.

Atmospheric muons are generated at the detector, applying a
weighted generation technique.
We generated a sample of $N_{total}= 2.5\cdot 10^7$ muons, in
the energy range 1 TeV $\div$ 1 PeV, with a generation
spectral index $X=3$. We also generated
$N_{total}= 4\cdot 10^7$ events in the range 100 GeV - 1 TeV,
with a generation spectral index $X=1$.
Muons are generated with an isotropic angular distribution.
The events are weighted to the
Okada parameterization \cite{Okada94}, taking into account the depth of the
Capo Passero site ($D=3500$ m) and the flux variation inside
the detector sensitive height ($h\approx900$ m).
According to the Okada parameterization, the expected number
of reconstructed muon events is about $4\cdot 10^8$ per year.
Our statistics cover only a few days. Considering that reconstructed events
have a flat distribution in Right Ascension (RA), we can project the
simulated events in a few degrees bin $\Delta$RA, centered in the source position.
So doing, we get statistics of atmospheric muons corresponding to a time
$\gtsim1$ year at all source declinations.

\begin{figure*}[t]
\begin{center}
\includegraphics[width=7.5cm]{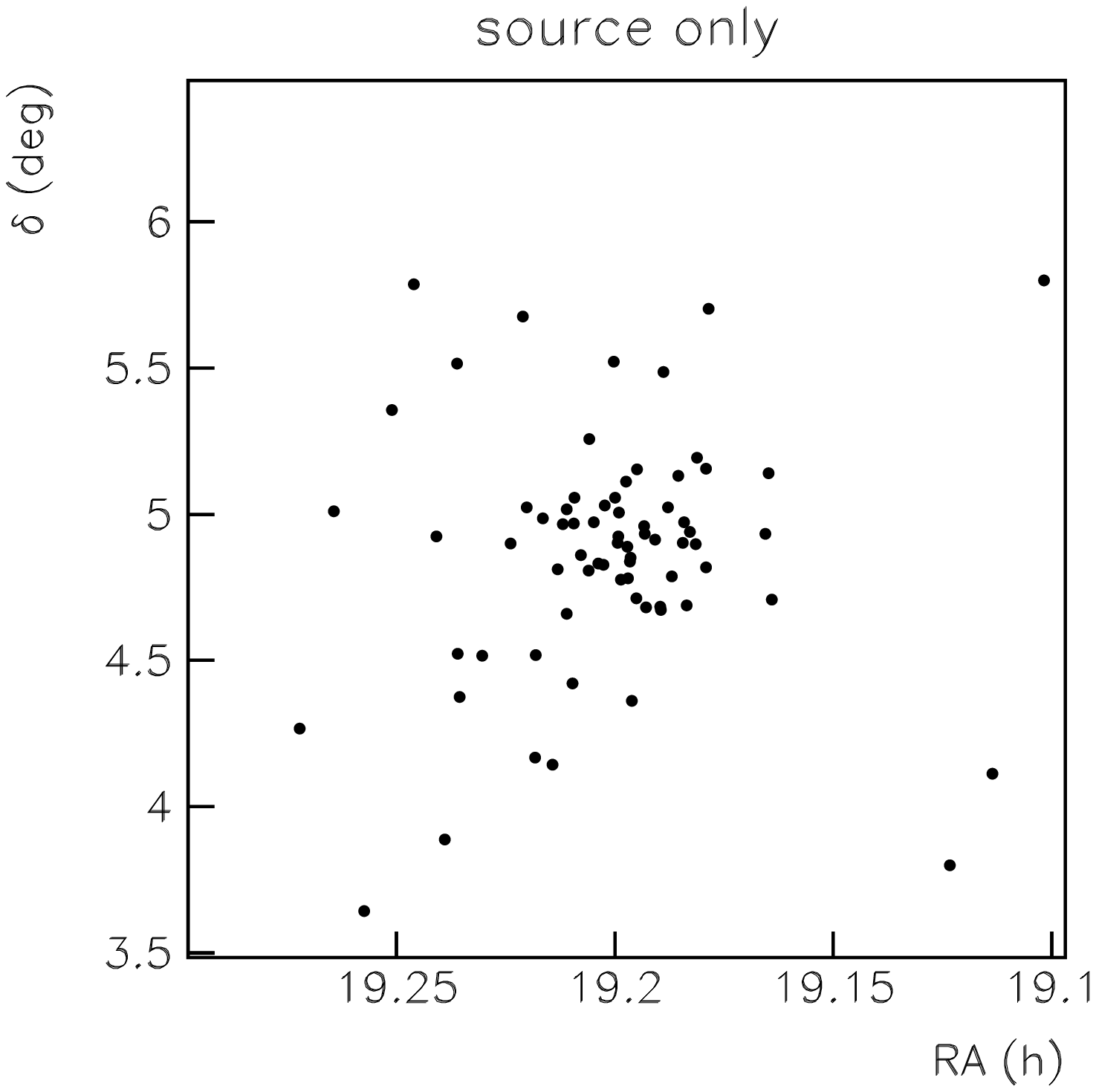}\includegraphics[width=7.5cm]{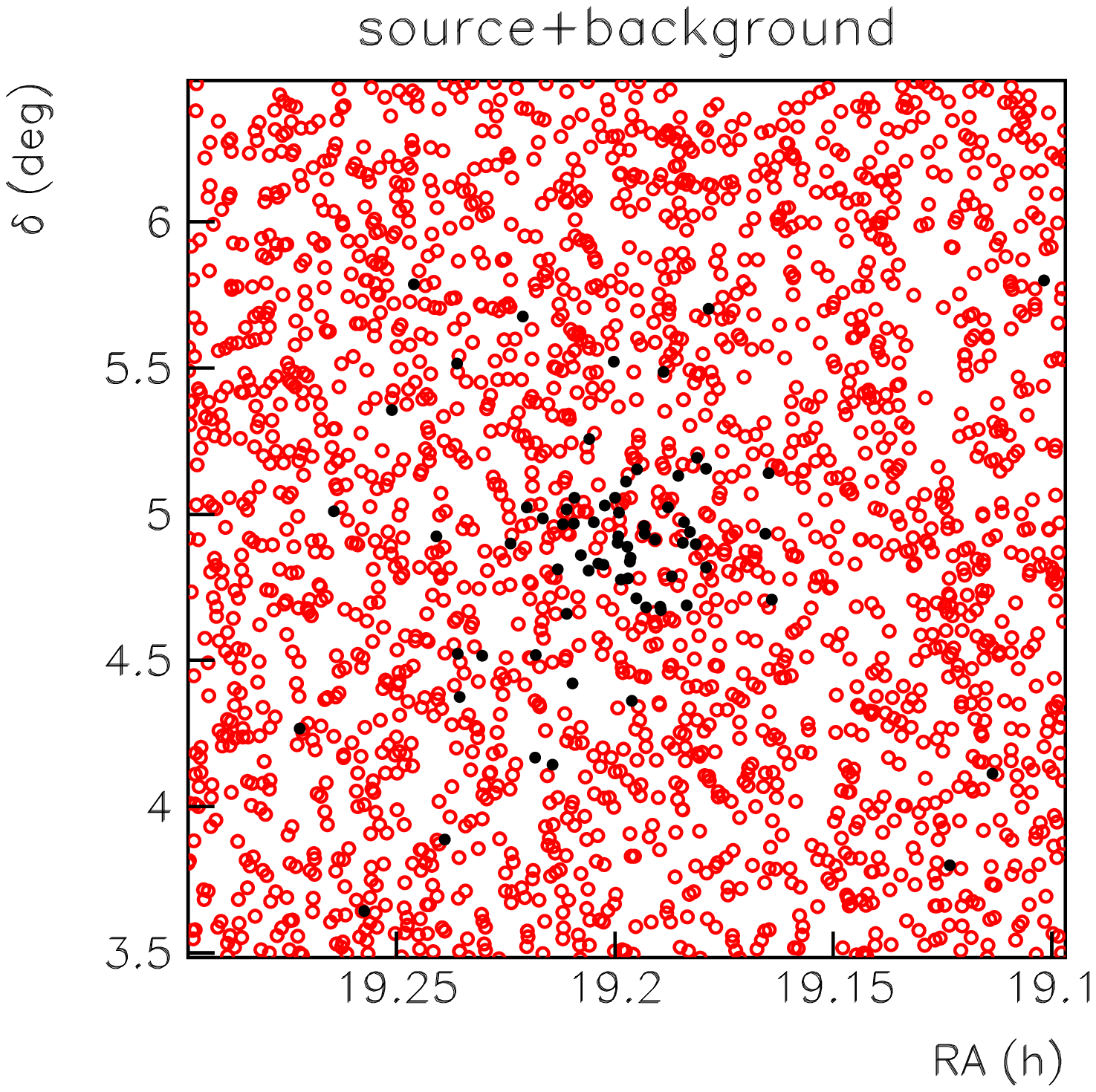}
\end{center}
\vspace*{-1.5cm}
\caption{Maps of the reconstructed events from SS433 (left panel) during an observation time of 1 year.
In the right panel background events are plotted together with the source events.
}
\label{fig:ss433}
\end{figure*}

At the reconstruction level, the fraction of background events is too high and exceeds
the expected signal also in the case of the most intense microquasars.
As an example we present the case of microquasars SS443
($\delta$=+04$^\circ$ 58' 58''.0, RA=19$^h$ 11$^m$ 49$^s$.6),
considering an observation time of 1 year
and an angular bin of $3^\circ \times 3^\circ$ around the
source position in the sky.
The result for SS433 is shown in
Fig. \ref{fig:ss433} in which are represented
$N_\mu=71$ events from the source and $N_\mu=2113$ from
atmospheric background (only up-ward going reconstructed tracks are considered).
Source and background events are selected with
a {\it hit-or-miss} procedure.
This procedure consists in comparing, for each
event, the weight $W_{event}$ with a number $R$ randomly generated
in the range $[0,1]$. If $R \leq W_{event}$ the event is accepted,
if not the event is rejected.

\section{SENSITIVITY TO MUON NEUTRINOS FROM MICROQUASARS}

\subsection{Calculation of the detector sensitivity}

The detector sensitivity was calculated according to the Feldman and Cousins
approach \cite{Feldman98}. The 90\% c.l. sensitivity to a neutrino flux coming from a point-like source
is given by
\begin{equation}
f_{\nu,90}=\frac{\overline{\mu}_{90}(b)}{N_\mu}f_{\nu,0},
\end{equation}
where $\overline{\mu}_{90}(b)$ is the 90\% c.l. average upper limit for an
expected background with known mean value $b$ and no true signal \cite{Feldman98}, $f_{\nu,0}$
is an arbitrary point source flux inducing a mean signal $N_\mu$.

\subsection{Selection criteria for background rejection}

The used reconstruction algorithm is a robust track fitting procedure based
on a maximization likelihood me\-thod \cite{antares_codes}.
In this work, we used, as a {\it goodness of fit criterion}, the variable:
\begin{equation}
\Lambda\equiv -\frac{\hbox{log}(\mathcal{L})}{N_{DOF}}+0.1(N_{comp}-1),
\end{equation}
where $\hbox{log}(\mathcal{L})/N_{DOF}$ is the log-likelihood
per degree of freedom ($N_{DOF}$) and $N_{comp}$ is the total number of compatible
solutions found by the reconstruction program. In particular, events are selected if
the variable $\Lambda$ is greater than a given value $\Lambda_{cut}$.
This quality cut is here applied together with other selection criteria as listed in the following:

\begin{itemize}

\item the number of hits $N_{fit}$, used to reconstruct the muon track,
must be greater than a given value $N_{fit}^{cut}$;

\item the muon must be reconstructed with $\vartheta_\mu^{rec}<\vartheta_\mu^{max}$,
in order to reject down-going events;

\item only events reconstructed in a circular sky region
centered in the source position and having a radius of
$r_{bin}$ are considered.

\end{itemize}

The optimal values of $\Lambda_{cut}$, $N_{fit}^{cut}$, $\vartheta_\mu^{max}$ and $r_{bin}$ are chosen to optimize
the detector sensitivity.

\subsection{Results}

The detector sensitivity was calculated for a livetime of 1 year,
simulating a neutrino flux with spectral index $\Gamma=2$
in the energy range 1 - 100 TeV and taking
into account both atmospheric neutrino and muon background.
The study was carried out for each microquasar,
since the sensitivity is a function
of the source astronomical declination.

\begin{figure}[th]
\begin{center}
\includegraphics[width=8.5cm]{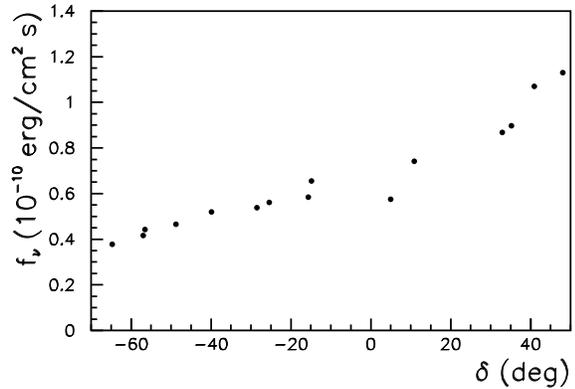}
\end{center}
\vspace*{-1.5cm}
\caption{NEMO-km$^3$ sensitivity to neutrinos from microquasars versus source declination,
for a livetime of 1 year. The worsening of the sensitivity with increasing declination is
due to the decrease of the source visibility.}
\label{fig:sensi-ene-flusso-delta-1years}
\end{figure}

In Tab. \ref{tab:sensitivity} we give the sensitivity for each
microquasar, and the corresponding $\vartheta_\mu^{max}$,
$\Lambda_{cut}$ and $r_{bin}$.
The dependence of the sensitivity on the source declination is
shown in Fig. \ref{fig:sensi-ene-flusso-delta-1years}.

\begin{table*}
\caption{Detector sensitivity to neutrinos from microquasars:
the sensitivity $f_\nu$ is calculated for an $\varepsilon_\nu^{-2}$
neutrino spectrum in the energy range 1 - 100 TeV, for a detector live
time of 1 year. The corresponding values of $\vartheta_\mu^{max}$, $\Lambda_{cut}$ and $r_{bin}$
and the source declination $\delta$ are also given.}
\label{tab:sensitivity}
\newcommand{\m}{\hphantom{$-$}}
\newcommand{\cc}[1]{\multicolumn{1}{c}{#1}}
\renewcommand{\tabcolsep}{2pc} 
\renewcommand{\arraystretch}{1.2} 
\begin{tabular}{@{}lcccc}
\hline
Steady Microquasars       &  $\vartheta_\mu^{max}$  (deg)   & $\Lambda_{cut}$ & $r_{bin}$ (deg) & $f_\nu$ (erg/cm$^2$ s)   \\
\hline
  LS 5039                 &  101    &   -7.3          & 0.9     & $6.5\cdot10^{-11}$    \\
  Scorpius X-1            &  104    &   -7.7          & 0.7 & $5.8\cdot10^{-11}$    \\
  SS433                   &  115    &   -8.0          & 0.8 & $5.7\cdot10^{-11}$    \\
  GX 339-4                &  ~96    &   -7.4          & 0.5 & $4.7\cdot10^{-11}$    \\
  Cygnus X-1              &  103    &   -7.5          & 0.7 & $9.0\cdot10^{-11}$    \\
\hline
Bursting Microquasars \\
\hline
  XTE J1748-288           &  102    &   -7.6          & 0.9     & $5.4\cdot10^{-11}$    \\
  Cygnus X-3              &  101    &   -7.3          & 0.8     & $1.1\cdot10^{-10}$    \\
  GRO J1655-40            &  101    &   -7.4          & 0.7     & $5.2\cdot10^{-11}$    \\
  GRS 1915+105            &  100    &   -7.4          & 0.8     & $7.4\cdot10^{-11}$    \\
  Circinus X-1            &  ~90    &   -7.3          & 0.9     & $4.2\cdot10^{-11}$    \\
  XTE J1550-564           &  ~90    &   -7.1          & 0.9     & $4.4\cdot10^{-11}$    \\
  V4641 Sgr               &  102    &   -7.4          & 0.9 & $5.6\cdot10^{-11}$    \\
  GS 1354-64              &  ~90    &   -7.5          & 1.0     & $3.8\cdot10^{-11}$    \\
  GRO J0422+32            &  103    &   -7.5          & 0.8 & $8.7\cdot10^{-11}$    \\
  XTE J1118+480           &  102    &   -7.5          & 0.7     & $1.1\cdot10^{-10}$    \\
\hline
\end{tabular}\\[2pt]
\end{table*}

In Tab. \ref{tab:sensitivity_aharonian} are summarized the detector sensitivities
for microquasar LS 5039, assuming neutrino fluxes with spectral indices $\Gamma=1.5$ and 2,
in the energy range 0.1 TeV and $\varepsilon_\nu^{max}=10$ and 100 TeV.

\begin{table*}
\caption{Detector sensitivity flux to neutrinos from LS 5039:
the sensitivity $f_\nu$ is calculated for a $\varepsilon_\nu^{-\Gamma}$
neutrino spectrum in the energy range from 0.1 TeV up to $\varepsilon_\nu^{max}$ and for a detector live
time of 1 year. The corresponding values of $\vartheta_\mu^{max}$, $\Lambda_{cut}$ and $r_{bin}$
are also given.}
\label{tab:sensitivity_aharonian}
\newcommand{\m}{\hphantom{$-$}}
\newcommand{\cc}[1]{\multicolumn{1}{c}{#1}}
\renewcommand{\tabcolsep}{2pc} 
\renewcommand{\arraystretch}{1.2} 
\begin{tabular}{@{}lccccc}
\hline
$\Gamma$ & $\varepsilon_\nu^{max}$ (TeV) &  $\vartheta_\mu^{max}$  (deg)   & $\Lambda_{cut}$ & $r_{bin}$ (deg) & $f_\nu$ (erg/cm$^2$ s)  \\
\hline
  1.5            & ~10      & 101      &  -7.3      & 1.0  & $1.5\cdot10^{-10}$   \\
  1.5            & 100      & 101      &  -7.3      & 0.9  & $5.2\cdot10^{-11}$    \\
  2.0            & ~10      & ~97      &  -7.4      & 1.0  & $2.7\cdot10^{-10}$   \\
  2.0            & 100      & 101      &  -7.3      & 0.9  & $9.6\cdot10^{-11}$    \\
\hline
\end{tabular}\\[2pt]
\end{table*}

\section{EXPECTED NUMBER OF EVENTS}

We applied the event selection cuts given in Tab. \ref{tab:sensitivity} and \ref{tab:sensitivity_aharonian} to determine
the expected number of events from theoretical models
considered in this paper.
In Tab. \ref{tab:N_mu_simulated} we give the number
of selected neutrino events from each microquasar, according to the
neutrino fluxes given by Distefano et al. \cite{Distefano02}.
In the same table we compare these results with the
atmospheric neutrino and muon background in 1 year of data taking.
In this analysis, it is assumed
that transient sources cause one burst per year, i.e. the
number of source events produced in the interval $\Delta t$ is
related to 1 year observation time.

The search for neutrino events in coincidence with microquasar
radio outbursts could restrict the analysis period to the flare
duration $\Delta t$, reducing the background. Such an analysis
technique, already used by AMANDA \cite{Amanda05}, can improve the
detector sensitivity to neutrinos from transient sources.
Referring to the bursts considered in Tab.
\ref{tab:N_mu_simulated} and integrating over the time interval
$\Delta t$, we expect about $10^{-3}$ background events per burst.

For the microquasar LS 5039 we considered the flux predicted by Aharonian et al. \cite{Aharonian05}.
The expected number of selected events is given in Tab. \ref{tab:N_mu_aharonian};
the comparison with the atmospheric background shows that an evidence could be
expected in a few years of data taking.

In Tab. \ref{tab:N_mu_bednarek}  we quote the expected number of events
from microquasar Cygnus X-3, according to the Bednarek model \cite{Bednarek05}.
Since the model predicts a neutrino flux with a spectral index close to 2
and since the main event contribution is in the energy range 1 - 100 TeV,
we used the same event selection parameters quoted in Tab. \ref{tab:sensitivity}.
Except the case of $\kappa=2.5$ and $\gamma_A^{max}=10^7$, the number of events
exceeds the expected background and the signal could be
detected in a few years of data taking.

\section{CONCLUSIONS}

The possibility to detect TeV neutrinos
from Galactic microquasars with the proposed NEMO-km$^3$ underwater
\v{C}erenkov neutrino telescope has been investigated.

A Monte Carlo was carried out to simulate the expected
neutrino-induced muon fluxes produced by point-like sources and by
atmospheric neutrinos. The expected atmospheric muon background
was also simulated.

We computed the detector sensitivity for each microquasar,
optimizing the event selection in order to reject the background.
Finally, we applied the event selection and calculated the number
of surviving events.

Our results show that,
assuming reasonable scenarios for TeV neutrino production,
the proposed NEMO telescope could identify microquasars in a few years of data taking,
with a discovery potential for at least few cases above the 5$\sigma$ level,
or strongly constrain the neutrino production models and the source parameters.

\end{document}